\documentclass{JHEP3}

\usepackage{amsfonts}
\usepackage{amsmath}
\usepackage{amssymb}
\usepackage{amsmath,amssymb,epsfig}
\numberwithin{equation}{section}
\usepackage{cite}

\newcommand{\be}{\begin{equation}}
\newcommand{\bea}{\begin{eqnarray}}
\newcommand{\eea}{\end{eqnarray}}
\newcommand{\ba}{\begin{array}}
\newcommand{\ea}{\end{array}}
\newcommand{\ee}{\end{equation}}

\newcommand{\bchi}{{\mbox{\boldmath $\chi$}}}

\title{First-order attractor flow equations  for supersymmetric black rings in N=2, D=5  supergravity}

\author{Yi-Xin
Chen and Yong-Qiang Wang \\
Zhejiang Institute of Modern Physics, Zhejiang University\\
 Hangzhou 310027, P. R. China\\
 E-mail: \email{yxchen@zimp.zju.edu.cn}, \email{wangyongqiangyueyuan@gmail.com}}

\abstract{In this paper we investigate the attractor mechanism in
the five dimensional low energy supergravity theory corresponding to
M-theory compactified on a Calabi-Yau threefold $CY_3$. Using very
special geometry, we derive the general first-order attractor flow
equations for BPS and non-BPS solutions in five-dimensional
Gibbons-Hawking spaces. Especially, considering the supersymmetric
solution, we obtain the first-order flow equations for
supersymmetric (multi)black rings.
We also solve the flow equations and discuss some properties of the
solutions of flow equations.}

\begin{document}

\section{Introduction}
The attractor mechanism in extremal black holes has been an
interesting subject over the past few years, which states that the
values of the moduli scalar fields at the horizon of the extremal
black holes are independent of the asymptotic values for the moduli
fields and are entirely determined by the quantized charges of the
black holes. It was initiated in the context of $N=2$ supergravity
theories in four dimensions \cite{1},
then extended to other supergravity theories and superstring
theories,
such as supersymmetric black holes with higher derivative
corrections , multi-center black holes and  spherically symmetric or
rotating black holes in higher dimensions\cite{2}. More recently,
more attention has been paid to non-supersymmetric cases \cite{3,
a}.

 In this mechanism, there exists a set of first-order differential equations, known as ``attractor flow
 equations'', which  describe the evolution of the spacetime
metric and the moduli fields in the background of extremal black
holes. In the literature, there are two methods to obtain these
equations: one can follow the method in\cite{1} , imposing the
preservation of supersymmetry,
the gravitino and gaugino variations vanish and lead to a set of
first-order flow
 equations about the metric and moduli fields . Another alternative
method is advised by Ferrara $et \, al. $ in \cite{a}, which is that
one can consider the attractor flow behavior as a result of an
extremization of the effective Lagrangian, rather than a
supersymmetric flow.

As is well known, the horizon topology of  black holes in five
dimensional spacetime is not unique.  The discovery of a new BH
phase made by Emparan and Reall\cite{4} : an asymptotically flat,
rotating black hole solution with horizon topology $S^1\times S^2$
and carrying angular momentum along the $S^1$, is called as black
ring. Several important developments are listed in \cite{5,6,7}. For
reviews, see\cite{8}.

It is interesting to understand the attractor mechanism in the
context of black ring. This mechanism has been addressed by Kraus
and Larsen in \cite{9}. In particular, by examining the the BPS
equations for black rings , they found the flow equations for
supersymmetric extremal black rings, an equation relating the flow
of the moduli to changes in the gauge field. The attractor mechanism
for the black rings determines the scalar values at the near-horizon
region via the magnetic dipole charges only. But, the flow equation
in \cite{9} is a second-order differential equation. It is obvious
to mention that do the first-order differential flow equations
analogous to the equations in \cite{1,3} can exit and, if so, how?

This is the aim of this article to further study the attractor
mechanism in five dimensional black rings. Recently, based on very
special geometry, Cardoso $et \, al. $ in \cite{10} proposed an
effective method for deriving  first-order flow equations for
rotating electrically charger extremal black holes
in five dimensions. Inspired by \cite{10}, we construct the general
first-order attractor flow equations for BPS and non-BPS solutions
in five dimensional Gibbons-Hawking spaces by making use of the
stationarity of actions. Especially, considering the configuration
of supersymmetric solution, we obtain the first-order attractor
equations for supersymmetric (multi)black rings, which take the form
analogous to that of a gradient flow of black holes in five
dimensions \cite{10, 18}. Because we do not analyze the attractor
flow equations following the method in \cite{9}, we are still
uncertain whether the second-order equation in \cite{9} can be
reduced to a set of first-order flow equations. However, when we
consider the condition of supersymmetric solutions,
the first-order flow equations which we obtain in five dimensional
Gibbons-Hawking spaces can reproduction a second-order equation the
flow equation derived in [9].
%
%
By integrating the first-order flow equations, we also find the
relation between the flow equations and electronic central charge
$Z_e$ corresponding to the graviphoton.

This paper is organized as follows. In the next section, we briefly
review the supersymmetric solutions of $N=2$ supergravity and
specialized to the case of (multi)black rings in the Gibbons-Hawking
spaces. Using very special geometry and the condition of stationary
of action, the generalized first-order attractor flow equations are
carried out in section \ref{flowequa}. In section \ref{some} we
present some properties of flow equations for supersymmetric black
rings and the example of the limit of black rings is given. The last
section is devoted to discussions.

\section{A brief review of supersymmetric solutions of $N=2$
supergravity}\label{intro} In this section we first present  a brief
review of the five dimensional low energy supergravity theory
corresponding to M-theory compactified on a Calabi-Yau threefold
$CY_3$. This model is usually studied in the context of real or very
special geometry. Relevant references  can be found in \cite{11, 12,
13,14}.
We also introduce the general supersymmetric solutions in
Gibbons-Hawking base space.
Further details, see\cite{6}.

\subsection{$N=2$ supergravity}
The bosonic part of $N=2$ $D=5$ ungauged supergravity coupled to
$n-1$ abelian vector multiplets with scalars $\phi^i, i= 1,...,n-1$,
is \bea \label{action}
 S &=& {1 \over 16 \pi G_{5}} \int \big(  R * 1 -G_{IJ} F^I
 \wedge *F^J -G_{IJ} dX^I \wedge
 * dX^J
\nonumber \\
  &-&{1 \over 6} C_{IJK} F^I \wedge F^J \wedge A^K \big) \, ,
\eea where the scalars $X^I= X^I(\phi^i)$, $I,J,K$ = $1 , ... ,n$,
obey the constraint : $\frac{1} { 6} C_{IJK} X^I X^J X^K=1$ and the
constants $C_{IJK}$ are symmetric on $IJK$.  It is useful to define
\bea
 X_I = {1 \over 6}C_{IJK} X^J X^K \, , \\
G_{IJ} = {9 \over 2} X_I X_J -{1 \over 2}C_{IJK} X^K \
\label{matrix}. \eea From the definitions, it follows that \be X^I
X_I= 1 \, ,\,\,\,\,\, X_A= \frac{2}{3}G_{AB} X^B \, ,\,\,\,\,\,X^A=
\frac{3}{2}G^{AB} X_B \, ,\ee and so \be X^I\partial_i X_I =
\partial_i X^I X_I=0\, .\ee By the definition (\ref{matrix}) one can
find \be
\partial_i X_I = -\frac{2}{3}G_{IJ}\partial_i X^J \, , \,\,\,
\partial_i X^I = -\frac{3}{2}G^{IJ}\partial_i X_J \,. \label{modulisca}\ee
The metric $g_{ij}$ on the scalar manifold is \be
g_{ij}=G_{AB}\partial_i X^A
\partial_j X^B \, ,\ee
where $\partial_i X^I= \frac{\partial X^I}{\partial \phi^i}$\,.
Combing these relations with the index structure we find  that \be
\label{vb} g^{ij}
\partial_i X^A \partial_j X^B = a(G^{AB} - b X^A X^B) \,,\ee with constant coefficients $a=1$ and $b= \frac{2}{3}$
.

We are interested in solutions preserving some supersymmetry.
Following the reference \cite{15,16},  supersymmetric solutions of
five-dimensional  supergravity imply the existence of  a
non-spacelike Killing vector field, and assuming that in a region
the Killing vector field $V=\partial/\partial t$  is time-like , the
$D=5$ metric is given by
\begin{equation}
 ds_5^2 = -f^2 (dt+\omega)^2 +f^{-1} ds_{\cal M}^2~,
\end{equation}
where $\cal M$ is a four-dimensional hyper-K\"{a}hler manifold, and
$f$ and $\omega$ are a scalar and  a 1-form on ${\cal M}$.
The field strengths $F^I$ can be written
\begin{equation}
F^I = d\left[fX^I(dt+\omega)\right] +\Theta^I~,
\end{equation}
where  $\Theta^I$ are closed 2-forms on the $\cal M$. The scalar
function $X_I$ and $f$, one-forms $\omega$ and $\Theta^I$ on $\cal
M$ are given by
\be (\Theta^I)^- = 0~, \,\,\, \triangle_{\cal M} (f^{-1} X_I)= {1
\over 6} C_{IJK} \Theta^J \cdot\Theta^K~,\,\,\, (d\omega)^+ =
-\frac{3}{2} f^{-1} X_I \Theta^I~, \label{bpsequa} \ee where
$\triangle_{\cal M}$ , the Laplacian, and the superscripts $\pm $ ,
self-dual and antiself-dual , are defined with respect to the base
$\cal M$, and for 2-forms $\alpha$ and $\beta$ on $\cal M$ we define
$\alpha \cdot \beta = \alpha^{mn } \beta_{mn}$, with indices raised
by the matrix $h^{mn}$ on $\cal M$. This equations are named as BPS
equations.

\subsection{Gibbons-Hawking base spaces and black ring solutions}
Now let us concentrate on the so-called Gibbons-Hawking base spaces.
In this paper, as the base space $\cal M$, we consider the
Gibbons-Hawking metric, which can then be written as \be ds_{\cal
M}^2 = H^{-1} (dx^5 + \chi)^2 +H \delta_{ij} dx^j dx^j \,, \ee where
$H$ is harmonic on the Euclid space $\mathbb{E}^3$, $\chi=\chi_i
dx^i$, $i, j = 1,\, 2, \, 3$, and $H$, $\chi$ are independent of
$x^5$ and can be solved explicitly, $\chi$ is determined by $\nabla
\times \bchi = \nabla H$. In this section $\nabla$ will be the
gradient and $\nabla^2$ will be the Laplacian on $\mathbb{E}^3$.

We introduce one-forms $\eta^I$, $\Theta^I = d\eta^I $. It is
convenient to set \bea \omega = \omega_5 (dx^5 + \chi)+ \hat{\omega}_4 \, ,\\
\eta^I =\eta^I_5 (dx^5 + \chi)+ \hat{\eta}_4^I \, ,\eea where
$\hat{\omega}_4=\omega_{4i} dx^i,~\hat{\eta}_4^I=\eta^I_{4i} dx^i$.
We can solve the BPS equation and obtain \cite{6}: \bea \nabla
\times
{\hat{\eta}^I_4} &=& -\nabla (H \eta^I_5 ) \, ,\label{addicon}\\
\nabla \times {\hat{\omega}_4} &=& H \nabla \omega_5 - \omega_5
\nabla H +3 \,H\big(f^{-1}X_I \big) \nabla \eta_5^I
\,,\label{formeta}\\f^{-1} X_I &=& {1 \over 24} H^{-1} C_{IPQ} K^P
K^Q +L_I, \eea where $\eta_5^I = \frac{1}{2} H^{-1} K^{I}$, and
$L_I,\,K^I$ are harmonic functions on $\mathbb{E}^3$.
Using the integrability condition of equation (\ref{formeta}), we
find the constraint \bea \nabla^2 \omega_5 = \nabla^2 \big( -{1
\over 48} H^{-2} C_{IPQ} K^I K^P K^Q -{3 \over 4} H^{-1} L_I K^I
\big)\, . \eea The solution of this equation is read as \be
\label{eqn:omvsol} \omega_5 = - {1 \over 48}H^{-2}C_{IPQ} K^I K^P
K^Q-{3 \over 4} H^{-1} L_I K^I +B \,,\ee where $B$ is another
harmonic function on $\mathbb{E}^3$. The general solution with
Gibbons-Hawking base is specified by $2n+2$ harmonic functions $H$,
$K^I$, $L_I$ and $B$ on $\mathbb{E}^3$. It is well known that $H$
determines the Gibbons-Hawking base, such as three examples of
Gibbons-Hawking metrics: flat space ($H = 1$ or $H = 1/|{\bf x}|$),
Taub-NUT space ($H = 1 + 2M/|{\bf x}|$) and the Eguchi-Hanson space
($H = 2M/|{\bf x}| + 2M/|{\bf x-x_0}|$) \cite{15}.
 It
is convenient to take the base space $\cal M$ to be flat space
$\mathbb{E}^4$ with metric  \be \label{hypkahl} ds^2_{\mathbb{E}^4}
= H^{-1} (d \psi+\chi)^2+ H(dr^2+r^2 \big[d \theta^2 + \sin^2 \theta
d \phi^2 \big])\, , \ee where $H=1/|{\bf x}|\equiv 1/r$ and $\chi =
\cos\theta d\phi$, which satisfies $\nabla\times {\bchi}=\nabla H$.
The range of the angular coordinates are $0<\theta<\pi$, $0<\phi<2
\pi$ and $0<\psi<4\pi$.

We are interested in the solutions of supersymmetric black ring
(multi-black rings). In \cite{6} the multi-black rings solutions is
given
by \bea \label{eqn:newgensol} K^I &=& \sum_{i=1}^M q^I{}_i h_i \,, \nonumber \\
L_I &=& \lambda_I  +{1 \over 24} \sum_{i=1}^M (Q_{Ii} - C_{IJK}
q^J{}_i q^K{}_i) h_i \,, \nonumber \\ B &=& \frac{3}{4}
\sum_{i=1}^M \lambda_I q^I{}_i - \frac{3}{4}
 \sum_{i=1}^M \lambda_I q^I{}_i |{\bf{x}}_i|h_i \,,
\eea where $h_i$ are harmonic functions in $\mathbb{E}^3$ centred at
${\bf x}_i$, $h_i=1/|{\bf x}-{\bf x}_i|$, and $Q_{Ii}$, $q^I_i$ and
$\lambda_I$ are constants.

\section{First-order flow equations in  Gibbons-Hawking spaces \label{sec:flow5dw5}
}\label{flowequa} In this section, we derive the general first-order
attractor flow equations for BPS and non-BPS solutions in five
dimensional Gibbons-Hawking spaces with the metric \eqref{hypkahl},
following the method of \cite{10}. To simplify the calculation , we
assume that all black rings  are sitting along the negative $z$-axis
of the three-dimensional space, ie, ${\bf x}_i$ located along the
$z$-axis , then, $K^I, L_I, B$ in (\ref{eqn:newgensol})only depend
on $r,\, \theta$.

Take the five-dimensional configuration as follows:
\begin{eqnarray}
     ds_5^2 &=& G_{MN} \, dx^M \, dx^N = - f^2 (r,\theta) \, (dt + w)^2 + f^{-1}(r,\theta) \, ds^2_{\mathbb{E}^4}
     \;,\nonumber \\
    A^I &=& \chi^I (r,\theta) \, (dt + w)+\eta^I \;,\nonumber \\
    w &=& w_5 (r,\theta) \, ( d\psi + \cos \theta \, d \phi) + w_4 (r,\theta)   \, d
    \phi\;,\nonumber \\
    \label{ee} \eta^I &=& \eta_5^{I}(r,\theta) \, (d \psi + \cos \theta \, d \phi)+ \eta_4^{I}(r,\theta)\, d
    \phi \;,
\end{eqnarray}
where $ds^2_{\mathbb{E}^4}$ is the metric (\ref{hypkahl}).
Substituting \eqref{ee} into the action \eqref{action}, we find that
the bosonic part of the five-dimensional $N=2$ supergravity action
can be expressed as
\begin{equation}
    \frac{8 \pi G_5}{V} \; S = S_1 + S_2 + S_3 +S_4\;,
\end{equation}
with $ V= 1/2$. Our goal is to find the first-order attractor flow
equations from this actions. We can proceed in the following steps.
Step one: The action $S$ is express in terms of squares of
first-order terms and total derivative terms. Then, step two:
Stationarity of action $S$ is imposed which implies all the
first-order terms vanish. After a tedious calculation, we obtain (we
refer to appendix  for some of the details)
\begin{eqnarray}
    S_1 &=& \frac12 \int dt \, dr \, d \theta \, d \phi \, d \psi\, \nonumber\\
    &&  \Bigl[\sin \theta \Bigl( -3 r^2 f^{-2} \partial_r^2 f - 2 r^2 g_{ij} { \partial_r\phi^i} { \partial_r\phi^j} + 2 r^2 f^{-2} G_{AB} { \partial_r\chi^A} \, { \partial_r\chi^B} \nonumber\\
    &&+ 2 \partial_r \left(r^2 f^{-1}  \partial_r f \right) \Bigr)+ \Bigl(  -3 \sin\theta f^{-2} \partial_\theta^2 f - 2 \sin\theta g_{ij} { \partial_\theta \phi^i} { \partial_\theta \phi^j} \nonumber\\
    &&+ 2 \sin\theta f^{-2} G_{AB} { \partial_\theta \chi^A} \, { \partial_\theta\chi^B} + 2 \partial_\theta \left(\sin\theta f^{-1}  \partial_\theta f \right) \Bigr)       \Bigr]\;,
\end{eqnarray}
This action is a function of the  spacetime metric $f$ and the
moduli $\phi$, including their derivatives. It can lead to the
Einstein equation and scalar equations of motion.
\begin{eqnarray}
    S_2 &=& \frac12 \int dt \, dr \, d \theta \, d \phi \, d \psi\, \sin \theta \nonumber\\
    && \Bigl[ - 2  \frac{f}{r}G_{AB}(\eta_5^A \eta_5^B  + \partial_\theta \eta_5^A\partial_\theta \eta_5^B -2 \csc \theta \,\eta_5^A\partial_\theta \eta_4^B \nonumber\\
    && + \csc^2\theta \, \partial_\theta \eta_4^{A}\partial_\theta \eta_4^{B}+ r^{2} \, \partial_r \eta_5^{A} \partial_r \eta_5^{B} + r^{2}\csc^2\theta \,\partial_r \eta_4^{A}\partial_r \eta_4^{B} ) \nonumber\\
    && + 4 \frac{s f }{\sin\theta} G_{AB}    \Bigl( \partial_r(\eta_5^B\cos\theta + \eta_4^B)\partial_\theta \eta_5^A -\partial_\theta(\eta_5^B\cos\theta + \eta_4^B)\partial_r \eta_5^A  \Bigr) \Bigr]
    \;.
\end{eqnarray}
This action is a function of the gauge potential $\eta^I$, and its
derivatives.
\begin{eqnarray}
    S_3 &=&  \int dt \, dr \, d \theta \, d \phi \, d \psi\,  \nonumber\\
    && \Bigl[ \frac{f \sin \theta }{2 r}\Bigl(  ( w_{5}^{2} + \partial_\theta^{2}w_{5} -2 \csc \theta \,w_{5}\partial_\theta w_{4} + \csc^2\theta \, \partial_\theta w_{4}^{2}+ r^{2} \, \partial_r w_{5}^{2} \nonumber\\
    && + r^{2}\csc^2\theta \,\partial_r w_{4}^{2}  ) (f^2 - 2 G_{AB}\chi^A \chi^B)  - 4 G_{AB} \chi^A (\partial_\theta \eta_5^{B} \, \partial_\theta w_{5} \nonumber\\
    && +  \csc^2\theta \, \partial_\theta \eta_4^{B} \, \partial_\theta w_{4}- \csc\theta w_{5} \, \partial_\theta \eta_4^{B} + \eta_5^B w_{5} + r^{2} \, \partial_r \eta_5^{B} \partial_r w_{5} \nonumber\\
    &&+ r^{2}\csc^2\theta \,\partial_r \eta_4^{B}\partial_r w_{4} )\Bigr)+\frac{1}{3V}  C_{ABC} \chi^A\chi^B\chi^C \Bigl((w_5 \sin\theta -\partial_\theta w_4)\partial_r w_5 \nonumber\\
    &&+ \partial_r w_4 \partial_\theta w_5 \Bigr ) -\frac{1}{2V}  C_{ABC}\chi^A\chi^B \Bigl(\partial_r w_5 \partial_\theta \eta_4^C -  \eta_5^C \sin\theta \partial_r w_5 \nonumber\\
    && - \partial_\theta w_5  \partial_r \eta_4^C + \partial_\theta w_4  \partial_r \eta_5^C   -\partial_r w_4  \partial_\theta \eta_5^C  -w_5\sin\theta\partial_r \eta_5^C               \Bigr)  \nonumber\\
    &&- 9 \frac{X_A X_B}{V^2}s f \Bigl( \partial_r(\eta_5^B\cos\theta + \eta_4^B)\partial_\theta \eta_5^A -\partial_\theta(\eta_5^B\cos\theta + \eta_4^B)\partial_r \eta_5^A          \Bigr)     \Bigr]
\end{eqnarray}
and
\begin{eqnarray}
    S_4 &=& \int dt \, dr \, d \theta \, d \phi \, d \psi\,  \nonumber\\
    && \frac{1}{3V}  C_{ABC} \Bigl[ \frac{1}{2}\partial_r\Bigl( \chi^A\chi^B \eta_5^{C}\partial_\theta(w_5 \cos\theta + w_4)- \chi^A\chi^B (\eta_5^C \cos\theta + \eta_4^C)\partial_\theta w_5 \nonumber\\
    && +2 \chi^A(\eta_5^B\partial_\theta \eta_4^C - \eta_5^B \eta_5^C \sin\theta - \partial_\theta \eta_5^B \eta_4^C          )       \Bigr)  \nonumber\\
    && - \frac{1}{2}\partial_\theta \Bigl( \chi^A\chi^B \eta_5^{C}\partial_r(w_5 \cos\theta + w_4)- \chi^A\chi^B (\eta_5^C \cos\theta + \eta_4^C)\partial_r w_5 \nonumber\\
    && +2 \chi^A(\eta_5^B\partial_r \eta_4^C  - \partial_r \eta_5^B \eta_4^C  )                   \Bigr)    \Bigr]   \,.             \label{bulkeva}
\end{eqnarray}
$S_3$  contains terms that are proportional to $\eta^I$ and $w$  and
derivatives thereof. $S_4$ is composed of the total derivative
terms, so ,we call $S_4$ as the boundary term.   $S_2$ and $S_3$
determine the  evolution of the gauge potential $A^I$ and $w$ . We
can also find that all four parts of the action can split into two
parts : one about coordinate $r$ and another about $\theta$.

The terms in $S_1$ can be written as:
\begin{equation}
S_1= S_1^{(a)}+S_1^{(b)} \,,
\end{equation}
\begin{eqnarray}
    S_1^{(a)} &=& \frac12 \int dt \, dr \, d \theta \, d \phi \, d \psi\, \sin \theta \left[-3 r^2 f^{-2} (\partial_r f)^2 - 2 r^2 g_{ij} {\partial_r \phi^i} {\partial_r \phi^j} \right.
    \nonumber\\
    && \qquad \left.
    + 2 r^{-2} f^{-2} G_{AB} \left( r^2 {\partial_r \chi^A} + f^2 G^{AC} V_C \right) \left( r^2 {\partial_r \chi^B} + f^2 G^{BD} V_D \right)  \right.
    \nonumber\\
    && \left.
    \qquad - 2 r^{-2} f^2 \,V_A G^{AB} V_B + 2 \partial_r \left(r^2 f^{-1} \partial_r f -2 V_A \, \chi^A \right) + 4 \chi^A \partial_r V_A \right]
    \;,
    \label{action1rew} \nonumber\\
    S_1^{(b)} &=& \frac12 \int dt \, dr \, d \theta \, d \phi \, d \psi\,  \left[-3 \sin\theta f^{-2} \partial_\theta^2 f - 2 \sin\theta g_{ij} { \partial_\theta \phi^i} { \partial_\theta \phi^j} \right.
    \nonumber\\
    && \qquad \left.
    + 2  \csc\theta \, f^{-2} G_{AB} \left( \sin\theta \, {\partial_\theta \chi^A} + f^2 G^{AC} U_C \right) \left( \sin\theta \, {\partial_\theta \chi^B} + f^2 G^{BD} U_D \right)\right.
    \nonumber\\
    && \qquad \left.
     - 2  \csc\theta \, f^2 \,U_A G^{AB} U_B
  + 2 \partial_\theta \left(\sin\theta f^{-1}  \partial_\theta f   - 2 U_A \, \chi^A \right) + 4 \chi^A \partial_\theta U_A \right]
  \;,
    \label{action1rew}
\end{eqnarray}
where  $V=V(r,\theta)$ and $U=U(r,\theta)$ are   scalar functions.
The term proportional to $V_A G^{AB} V_B$ , with the definition of
\eqref{vb}, can be written as
\begin{equation}
    V_A G^{AB} V_B = \frac23 V_A \, X^A X^B V_B+ g^{ij} \,V_A \partial_i X^A \, \partial_j X^B V_B \;, \label{effpotz}
\end{equation}
we can also obtain the similar result about the term $U_A G^{AB}
U_B$. Then, using these relations, we obtain
\begin{eqnarray}
    \label{stationarySBrS1} S_1^{a} &=& \frac12 \int dt \, dr \, d \theta \, d \phi \, d \psi\, \sin \theta \biggl[-3 \tau^2 f^2 \left( \partial_{\tau} f^{-1} - \frac23 V_A X^A \right)^2 \nonumber\\
    && - 2 \tau^2 g_{ij} \left( \partial_{\tau} \phi^i + f g^{il} V_A  \partial_l X^A \right) \left( \partial_{\tau} \phi^j + f g^{jk} V_B \partial_k X^B \right) \nonumber\\
    && \left.
    + 2 \tau^2 f^{-2} G_{AB} \left(\partial_{\tau} {\chi^A} - f^2 G^{AC} V_C \right) \left( \partial_{\tau} {\chi^B} - f^2 G^{BD} V_D \right) \right.
    \nonumber\\
    && + 2 \partial_r \left(r^2 f^{-1} f' -2 V_A \, \chi^A - 2 f \,  V_A X^A \right) + 4 \chi^A \partial_r V_A +4 f X^A \partial_r V_A \biggr] \;,
\end{eqnarray}
where $\tau = \frac{1}{r}$ \;.
\begin{eqnarray}
    \label{stationarySBrS2} S_1^{b} &=& \frac12 \int dt \, dr \, d \theta \, d \phi \, d \psi\,  \biggl[-3 \csc\theta \, f^2 \left(\sin\theta \, f^{-2} \partial_{\theta} f - \frac23 U_A X^A \right)^2 \nonumber\\
    && - 2 \csc\theta \, g_{ij} \left(\sin\theta \, \partial_{\theta} \phi^i - f g^{il} U_A  \partial_l X^A \right) \left(\sin\theta \, \partial_{\theta} \phi^j - f g^{jk} U_B \partial_k X^B \right) \nonumber\\
    && \left.
    + 2  \csc\theta \, f^{-2} G_{AB} \left( \sin\theta \, {\partial_\theta \chi^A} + f^2 G^{AC} U_C \right) \left( \sin\theta \, {\partial_\theta \chi^B} + f^2 G^{BD} U_D \right) \right.
    \nonumber\\
    && + 2 \partial_\theta \left(\sin\theta f^{-1}  \partial_\theta f   - 2 U_A \, \chi^A  - 2 f \,  U_A X^A \right)
    \nonumber\\
    && + 4 \chi^A \partial_\theta U_A +4 f X^A \partial_\theta U_A \biggr]\,.
\end{eqnarray}
When requiring stationarity of $S_1$ with respect to variations of
the fields, the last two terms in \eqref{stationarySBrS1} and
\eqref{stationarySBrS2} vanish.  Thus, up to a total derivative
term, $S^a_1$ is expressed in terms of squares of first-order flow
equations which  result in \bea
    \label{Br1flow}
        \partial_{\tau} f^{-1} &=& \frac23 V_A X^A \;, \label{flowfi} \nonumber\\
        \partial_{\tau} {\chi^A} &=& f^2 G^{AC} V_C  \;, \label{flowchi} \nonumber\\
        \partial_{\tau} \phi^i &=& - f g^{il} V_A  \partial_l X^A \;.
        \label{flowphi}
\eea In the same result from  $S^b_1$, we can obtain : \bea
    \label{Br2flow}
        \sin\theta \, f^{-2} \partial_{\theta} f  &=& \frac23 U_A X^A \;, \label{flowfi}\nonumber\\
        \sin\theta \, {\partial_\theta \chi^A} &=& - f^2 G^{AC} U_C  \;, \label{flowchi} \nonumber\\
        \sin\theta \, \partial_{\theta} \phi^i &=&   f g^{il} U_A  \partial_l X^A\;.
        \label{flowphi}
\eea
Eqs. (\ref{Br1flow}) and (\ref{Br2flow}) describe the evolution of
the spacetime metric and the moduli fields in the background of five
dimensional Gibbons-Hawking base space, which are analogous to the
flow equations for supersymmetric black holes in asymptotically flat
spacetime in five dimensions derived in \cite{10,18}. It is worth
pointing out that these first-order equations were derived without
using supersymmetry, therefore, attractor flow equations
(\ref{Br1flow}) and (\ref{Br2flow}) can including the supersymmetric
and non-supersymmetric case. In next section, we will rewrite above
two set of equations into one compact form.

Also, we rewrite $S_2$ as a sum of squares, as follows.
\begin{eqnarray}
S_2 &=& -  \int dt \, dr \, d \theta \, d \phi \, d \psi\, r^{-1} f \sin \theta \, G_{AB} \nonumber\\
    && \Bigl[ (r \, \partial_r \eta_5^{A} - s\,(\eta_5^A  - \csc\theta \, \partial_\theta \eta_4^{A}) ( r \, \partial_r \eta_5^{B}- s\,( \eta_5^B - \csc\theta \, \partial_\theta \eta_4^{B})\nonumber\\
    &&  + (\partial_\theta \eta_5^A -s\, r\csc\theta \,\partial_r \eta_4^{A} )(\partial_\theta \eta_5^B  -s\, r \csc\theta \,\partial_r \eta_4^{B} )  \Bigr]
    \;.
\end{eqnarray}
thus, we obtain one additional  first-order  equations about
$\eta^I$ following the stationarity of $S_2$:
\begin{subequations}
    \label{Br1flow5drot}
    \begin{align}
        r \, \partial_r \eta_5^{A} &= s\,(\eta_5^A  - \csc\theta \, \partial_\theta \eta_4^{A})  \;, \label{rot11}\\
        \partial_\theta \eta_5^A &=s\, r\csc\theta \,\partial_r \eta_4^{A} \;.
    \end{align}
\end{subequations}
Also, we rewrite the Eqs.(\ref{Br1flow5drot}) in compact form: \be
\nabla \times {\hat{\eta}^I_4} =-\nabla (H \eta^I_5 )
\label{addicon1} \,, \ee with $H= r^{-1}$. We reproduces the
Eqs.(\ref{addicon}) precisely.

Next, we rewrite $S_3$ as a sum of squares, as follows. Using the
definition in\cite{10}
\begin{equation}
    \label{psitilde} \chi^A =  - s f \,  X^A \,,
\end{equation}
with $s=1$, Then, we obtain for $S_3$,
\begin{eqnarray}
    S_3 &=& - \int dt \, dr \, d \theta \, d \phi \, d \psi\, r^{-1}f^3 \sin \theta  \nonumber\\
    && \Bigl[ \Bigl( r \, \partial_r w_{5} + s( w_{5}  - \csc\theta \, \partial_\theta w_{4})+3 \,s\, r f^{-1} X_A \partial_r \eta_5^A \Bigr) \Bigl( r \, \partial_r w_{5} \nonumber\\
    &&  + s( w_{5}  - \csc\theta \, \partial_\theta w_{4})+3\,  f^{-1} X_A(\eta_5^A - \csc\theta \partial_\theta \eta_4^A) \Bigr)\nonumber\\
    && + (\partial_\theta w_{5} + s\,r \csc \theta \,\partial_r w_{4} + 3\,s\, f^{-1} X_A \partial_\theta \eta_5^{A})(\partial_\theta w_{5} + s\,r \csc \theta \,\partial_r w_{4} \nonumber\\
    &&+ 3 \, r \csc\theta f^{-1} X_A \partial_r \eta_4^{A} )\Bigr] \;.
\end{eqnarray}
Then, another additional first-order flow equations following from
the stationarity of $S_3$ are
\begin{subequations}
    \label{BPSflow5drot}
    \begin{align}
        r \, \partial_r w_{5} + s( w_{5}  - \csc\theta \, \partial_\theta w_{4})&=-3 \,s\, r f^{-1} X_A \partial_r \eta_5^A  \;, \label{rot1}\\
         r \, \partial_r w_{5} + s( w_{5}  - \csc\theta \, \partial_\theta w_{4})&=-3\,  f^{-1} X_A(\eta_5^A - \csc\theta \partial_\theta \eta_4^A)  \;, \label{rot2}\\
        \partial_\theta w_{5} + s\,r \csc \theta \,\partial_r w_{4} &=- 3\,s\, f^{-1} X_A \partial_\theta \eta_5^{A} \;, \label{rot3}\\
        \partial_\theta w_{5} + s\,r \csc \theta \,\partial_r w_{4}&=- 3 \, r \csc\theta f^{-1} X_A \partial_r \eta_4^{A}  \;.
    \end{align}
\end{subequations}
Considering Eqs. (\ref{Br1flow5drot}), we can obtain :\bea
\label{eqn:selfom} \nabla \times {\hat{\omega}_4} &=& H\nabla
\omega_5 - \omega_5 \nabla H +3 \,H\big(f^{-1}X_A \big) \nabla
\eta_5^A \label{addicon2} \;, \eea with $H= r^{-1}$. It is obvious
to see that the gauge fields $A^I$ and one-form $\omega$ are subject
to the constraint of equations (\ref{addicon1}) and
(\ref{addicon2}).

So far we have discussed the bosonic part of the supergravity action
in five dimensional Gibbons-Hawking base.
Assuming the stationary of action, we obtain the general first-order
flow equations (\ref{Br1flow}) and (\ref{Br2flow}), which describe
the evolution of the metric and the moduli fields in the background
of general five dimensional solution. Meanwhile, the constraint
(\ref{addicon1}) and (\ref{addicon2}) can be obtain. Since the
present discussion is that it does not rely on supersymmetry, the
above conclusion can include both BPS and non-BPS solutions.

\section{Some properties of attractor flow equations for supersymmetric black rings}\label{some}
In order to solve the flow equations (\ref{Br1flow}) and
(\ref{Br2flow}), we need rewrite two equations
into one compact form. First, we introduce a anszta : \be \label{xx}
f^{-1} X_I = \zeta_I \;,\ee where $\zeta_I=\zeta_I(r, \theta)$.  We
take the gradient of  this equation with respect to the base space
$\mathbb{E}^3$ and obtain :
\begin{equation}\label{eq}
f^{-1}\nabla X_I +  X_I\nabla f^{-1}  =\nabla \zeta_I\;.
\end{equation}
Using the relation $X_I X^I =1$, we also have: $X^I\nabla X_I=\nabla
X^I X_I =0$, and consequently, we write Eqs.(\ref{eq}) as:
\begin{equation}\label{eqn:scalsol}
\nabla f^{-1}  =X^I \nabla\zeta_I \;.
\end{equation}
Comparing  Eqs.(\ref{eq}) with Eqs. (\ref{Br1flow}) and
(\ref{Br2flow}), we obtain
\begin{equation}\label{eqn:scalsol}
 V_I =- \frac{3}{2} r^2\,\partial_r\,\zeta_I \,, \,\,\,\,\,\,\,\,\,\, U_I =- \frac{3}{2} \sin\theta\,\partial_\theta\,\zeta_I \,.
\end{equation}
So, we combine Eqs. (\ref{Br1flow}) and (\ref{Br2flow}) into
\begin{subequations}
    \label{BPSflow5d1}
    \begin{align}
        \nabla f^{-1}  &=X^I \nabla\zeta_I \;, \label{flowequaf1} \\
        \nabla  {\chi^A} &=\frac{3}{2} f^2 G^{AI} \nabla\zeta_I  \;, \label{flowequaf2} \\
        \nabla \phi^i &= - \frac{3}{2} f g^{il} \nabla\zeta_I \,  \partial_l X^I \;.
        \label{flowequaf3}
    \end{align}
\end{subequations}
These equations which are called the first-order attractor flow
equations are one of main results in this paper. The flow equations
 describe the evolution of the spacetime metric and the moduli.
Remarkably, we can observe that equations (\ref{BPSflow5d1}) take
the form analogous to that of a gradient flow of black holes in five
dimensions derived in \cite{10,18} .

So far we obtain the first-order flow equations in addition the
constraint Eqs. (\ref{addicon1}) and (\ref{addicon2}) for the
general solutions following the stationarity of actions $S$. We note
that in the above consideration we have not considered the
supersymmetric properties of the solutions, therefore the solution
would be BPS or non-BPS. When including the supersymmetry, we get
$f^{-1} X_I =\zeta_I= {1 \over 24} H^{-1} C_{IPQ} K^P K^Q +L_I$ by
solving the BPS equations (\ref{bpsequa}). So, the flow equations
(\ref{BPSflow5d1}) are solved
\begin{subequations}
    \label{BPSflow5}
    \begin{align}
        \ f^{-1} &=( {1 \over 24} H^{-1} C_{IPQ} K^P K^Q +L_I) X^I \;, \label{flowfi} \\
        \chi^A &= - s\, f X^A  \;, \label{flowchi} \\
        f^{-1}X_I &=   {1 \over 24} H^{-1} C_{IPQ} K^P K^Q +L_I\;,
        \label{flowphi}
    \end{align}
\end{subequations}
where $H^{-1}= r$, and $L_I,\,K^I$ are harmonic functions on
$\mathbb{E}^3$ and are given by Eqs.(\ref{eqn:newgensol}). The same
solutions of supersymmetric (multi)black rings have been obtain in
\cite{6, 12,13}.

We would like to emphasize that when considering the supersymmetric
configuration (\ref{bpsequa}), the flow equations (\ref{BPSflow5d1})
have the following properties :

(i) Integrating Eq.(\ref{flowequaf1}) on base space $\mathbb{E}^3$,
we can introduce a term $Z_e(V)$ as

\be Z_e(V)  = {3 \over 4\pi^2}  \int_{\partial V} \!
d\overrightarrow{\bf S}\,\cdot\,  \nabla f^{-1} ~,
\label{somepro1}\ee where $\partial V$ is a closed hypersurface in
base space $\mathbb{E}^3$. Using the outward pointing unit normal
vector ${\bf n}$ , we get \bea Z_e(V) &=& {3 \over 4\pi^2}
\int_{\partial V} \! dS\, f^{-2} {n}^m
\partial_m f ~,\nonumber\\
 &=& {1 \over 2\pi^2}  \int_{\partial V} \! dS\, f^{-1} X^I
{n}^mE_{ mI}~,\eea where \be E_{ mI}\equiv G_{IJ}F^J_{m\hat{t}}
,\,\,\, F^I_{m{\hat t}}=f^{-1}\partial_m(f X^I)\;.\label{defin}\ee
It is obvious to show that the definition (\ref{somepro1}) agrees
precisely with the electronic central charge in \cite{9}{\footnote{
Note that  we choose the three-dimensional submanifold of
four-dimensional hyper-K\"{a}hler manifold $\cal M$ in this section,
which is the only difference comparing with the choice in
\cite{9}.}}. We can consider that $Z_e$ is the electric charge
corresponding to the graviphoton.

 (ii)There is another form of the attractor formula that is cast entirely
in terms of the moduli space. To derive it, we multiply the term
$\partial_i X^I$ on both sides of (\ref{flowequaf3}), we can write
the result as \be \nabla X^I = - \frac{3}{2} f g^{il} \nabla\zeta_J
\,
\partial_l X^J \, \partial_i X^I. \ee Using the relation (\ref{modulisca}), we
get \be \nabla X^I = - \frac{3}{2} f G^{IJ} \nabla\zeta_K D_J X^K\,
, \ee where the covariant derivative is defined as $D_I=
\partial_I - \frac{1}{3} X_I$.

(iii) In condition of supersymmetric solutions, we take the
divergence of Eq. (\ref{flowequaf1}) and obtain: \bea
\nabla \cdot \nabla f^{-1}  &=& \nabla \cdot( X^I \nabla\zeta_I )\;\nonumber\\
&=& \nabla X^I \cdot \nabla\zeta_I +  X^I \nabla \cdot
\nabla\zeta_I\;\label{pro1}\nonumber\\
&=& -\frac{2}{3}f^{-1} G_{IJ} \nabla X^I \cdot \nabla X^J  + {1
\over 6} C_{IJK}X^I \Theta^J \cdot \Theta^K \;,
  \eea
where we used (\ref{modulisca}) and second  BPS equations
(\ref{bpsequa}) to arrive at the third line. We can rewrite
(\ref{pro1}) as \be \nabla^m (f^{-1} X^I E_{mI})=f^{-1} G_{IJ}
\nabla X^I \cdot \nabla X^J  - {1 \over 4} C_{IJK}X^I \Theta^J \cdot
\Theta^K \;,\ee where  $E_{mI}$ take the definition (\ref{defin}).
This equation agrees precisely with flow equations obtain  by Kraus
and Larsen in \cite{9}.

In the rest of this section we give a special case as an example. We
take the compactification manifold to be $T^6$, In this case
$C_{IJK} = 1 $ if $(IJK)$ is a permutation of $(123)$, and $C_{IJK}
= 0$ otherwise. The metric $G_{IJ}$ is \be G_{IJ}= \frac{1}{2}{\rm
diag}((X^1)^{-2},(X^2)^{-2},(X^3)^{-2})\;.\ee When M=1 and $x_1=0$
in (\ref{eqn:newgensol}), i.e. all of the harmonic functions in
$\mathbb{E}^3$ are centred at the origin. With the metric
(\ref{hypkahl}) , we find $\chi^I=H_I^{-1} = \lambda_I  + Q_I/r$,
and the gauge potential
\be \label{hju} A^I = H_I^{-1}(d t + \omega), \,\,\, \omega_5=
-\frac{J}{2 r},\,\,\, \omega_4=0 \,, \ee where $J=\frac{1}{2}(q_1
Q_1 + q_2 Q_2 + q_3 Q_3 - q_1 q_2 q_3) $. This is the BMPV black
hole with three independent charges $Q_i$ and angular momenta
$J_\phi  = J_\psi = J $ \cite{19}. Inserting (\ref{hju}) into  the
flow equations (\ref{BPSflow5d1}), we can obtain
\begin{subequations}
    \label{BPSflow5drot}
    \begin{align}
       \partial_\tau f^{-1}  &=Z_e \;, \label{flowfi11} \\
       \partial_\tau \phi^i &= - \frac{3}{2} f g^{il}   \partial_l Z_e
       \;,
       \label{flowphi}
    \end{align}
\end{subequations}
with $Z_e=X^I Q_I $. We can  use  an equivalent way to find  the
attractor point of these equations, which is that one can find the
fixed values of the moduli by extremizing the central charge $Z_e$.
Extremizing the central charge with respect to the fixed moduli
means that we impose $\partial_i Z_e = 0 $. We shall assume that the
charges $Q_I$ are chosen such that at the attractor point, $ Z_e=
Z_e^\ast\neq 0$. Thus, equation (\ref{flowfi11}) may be easily
integrated near the horizon, \be f^{-1} \sim Z_e^\ast/r \;.\ee The
Bekenstein-Hawking entropy is one quarter of the horizon area, \be
S_{BH}=  2\pi \lim_{r\to 0} \sqrt{(r\,f^{-1})^3 - J^2} = 2 \pi
\sqrt{(Z_e^\ast)^3 - J^2}\;. \ee We reproduce the BMPV black hole
entropy in five dimensions which have been obtained in \cite{19}.

\section{Conclusion}
In this paper, by the use of the stationarity of actions , we have
obtained the first-order attractor flow equations for the general
solutions of motion equations for $N=2$  supergravity in five
dimensional Gibbons-Hawking space. Meanwhile, we also get the
constraint (\ref{addicon1}) and (\ref{addicon2}) which determine the
gauge field $A^I$ and one-form $\omega$. Furthermore, when
considering the supersymmetry  we obtain the first-order flow
equations for supersymmetric
(multi)black rings. It is also showed that the supersymmetric
solution with Gibbons-Hawking base is specified by $2n+2$ harmonic
functions $H$, $K^I$, $L_I$ and $B$ on the flat space
$\mathbb{E}^3$. Using the very special geometry,  We analyze the
equation (\ref{flowequaf1}) in first-order flow equations and find
that the integrate of r.h.s. of (\ref{flowequaf1}) agrees
precisely with the electronic central charge $Z_e$ in \cite{9}.
Moreover, taking the divergence of (\ref{flowequaf3}), we can
reproduce the second-order flow equation which have been obtained
by Kraus and Larsen in \cite{9}. A particular case, BMPV black
hole which is the limits of the supersymmetric black ring
solution, is presented in the last.

\section*{Acknowledgement}

We would like to thank C. Cao, Q. J. Cao, Y. J. Du, J. L. Li, Q. Ma
and Y. Xiao  for useful discussions. The work is supported in part
by the NNSF of China Grant No. 90503009, No. 10775116, and 973
Program Grant No. 2005CB724508.

\appendix

\section{Evaluating the action in five dimensions \label{bulkaction5}}

The square root of the determinant of the metric
(\ref{ee}),(\ref{hypkahl}) is
\begin{equation}
    \sqrt{-G} = \frac{ r  \sin\theta}{f} \;.
\end{equation}
The inverse metric reads
\begin{equation}
    G^{MN} =
    \begin{pmatrix}
        -1/f^2 + f\left(\frac{ w_5^2}{r} + \frac{w_4^2}{r \sin^2\theta}\right) & 0 & 0 & -\frac{f w_4}{r }\frac{1}{\sin^2\theta} & f\left(-\frac{ w_5}{r} + \frac{  w_4}{r }\frac{\cos \theta}{\sin^2\theta}\right) \\
        0 & r f & 0 & 0 & 0 \\
        0 & 0 & f/r & 0 & 0 \\
        -\frac{f w_4}{r}\frac{1}{\sin^2\theta} & 0 & 0 & \frac{f}{r}\frac{1}{\sin^2\theta} & -\frac{f}{r}\frac{\cos\theta}{\sin^2\theta} \\
        f\left(-\frac{  w_5}{r} + \frac{  w_4}{r }\frac{\cos \theta}{\sin^2\theta}\right) & 0 & 0 & -\frac{ f}{r}\frac{\cos\theta}{\sin^2\theta} & f\left(\frac{1}{r} + \frac{1}{r}\frac{\cos^2\theta}{\sin^2\theta}\right)
    \end{pmatrix}
\end{equation}
and the Ricci scalar is
\begin{equation}
    \label{Ricci}
    \begin{split}
        R = \frac{1}{2 r^2 f }\biggl[ &   f^{5}(w_{5}^{2} + \partial_\theta^{2}w_{5} -2 \csc \theta \,w_{5}\partial_\theta w_{4} + \csc^2\theta \, \partial_\theta w_{4}^{2} \\
        &+ r^{2} \, \partial_r w_{5}^{2} + r^{2}\csc^2\theta \,\partial_r w_{4}^{2} ) -5 r (\partial_\theta^{2}f + r^{2} \partial_{r}^{2}f )\\
        & + 2 r f (\cot\theta \,\partial_{\theta}f +\partial_{\theta\theta} f +2 r \partial_{r} f + r^2 \partial_{rr}f  ) \biggr]
        .
    \end{split}
    \raisetag{5ex}
\end{equation}

Inserting the ansatz
\begin{eqnarray}
    A^I &=& \chi^I (r,\theta) \, (dt + w)+\eta^I ,
\end{eqnarray}
into the gauge kinetic term in (\ref{action}) yields
\begin{equation}
    \begin{split}
        -\frac{1}{2}\sqrt{-G}\, G_{AB} F^A_{MN} F^{BMN} = &-\frac{\sin\theta}{2 r f^2} G_{AB} \Bigl[2 f^3\Bigl(\eta_5^A \eta_5^B  + \partial_\theta \eta_5^A\partial_\theta \eta_5^B -2 \csc \theta \,\eta_5^A\partial_\theta \eta_4^B \\
        &+ \csc^2\theta \, \partial_\theta \eta_4^{A}\partial_\theta \eta_4^{B}+ r^{2} \, \partial_r \eta_5^{A} \partial_r \eta_5^{B} + r^{2}\csc^2\theta \,\partial_r \eta_4^{A}\partial_r \eta_4^{B}  \\
        &+ 2 \chi^A(\partial_\theta \eta_5^{B} \, \partial_\theta w_{5} -  \csc\theta \eta_5^{B} \, \partial_\theta w_{4} +  \csc^2\theta \, \partial_\theta \eta_4^{B} \, \partial_\theta w_{4}  \\
        &- \csc\theta w_{5} \, \partial_\theta \eta_4^{B} + \eta_5^B w_{5} + r^{2} \, \partial_r \eta_5^{B} \partial_r w_{5} \\
        &+ r^{2}\csc^2\theta \,\partial_r \eta_4^{B}\partial_r w_{4} ) + \chi^A \chi^B ( w_{5}^2  + \partial_\theta w_{5}^2   \\
        &-2 \csc \theta \,w_{5}\partial_\theta w_{4}+ \csc^2\theta \, \partial_\theta w_{4}^2+ r^{2} \, \partial_r w_{5}^2 + r^{2}\csc^2\theta \,\partial_r w_{4}^2                               ) \Bigr) \\
        &-2 r  ( \partial_\theta\chi^A \partial_\theta\chi^B  +   r^2 \partial_r \chi^A \partial_r \chi^B) \Bigr] \;.
    \end{split}
\end{equation}
The Chern--Simons term in (\ref{action}) evaluates to
\begin{eqnarray}\label{app1}
    - \frac{1}{6V} C_{ABC} \, F^A \wedge F^B \wedge A^C &=& \frac{1}{3V}  C_{ABC}\Bigl[\chi^A\chi^B\chi^C \Bigl((w_5 \sin\theta -\partial_\theta w_4)\partial_r w_5 + \partial_r w_4 \partial_\theta w_5 \Bigr )  \nonumber\\
    && -\frac{3}{2}\chi^A\chi^B \Bigl(\partial_r w_5 \partial_\theta \eta_4^C -  \eta_5^C \sin\theta \partial_r w_5 - \partial_\theta w_5  \partial_r \eta_4^C                           \nonumber\\
    && + \partial_\theta w_4  \partial_r \eta_5^C  -\partial_r w_4  \partial_\theta \eta_5^C  -w_5\sin\theta\partial_r \eta_5^C               \Bigr)  \nonumber\\
    && +3 \chi^A \Bigl( \partial_r(\eta_5^B\cos\theta + \eta_4^B)\partial_\theta \eta_5^C -\partial_\theta(\eta_5^B\cos\theta + \eta_4^B)\partial_r \eta_5^C          \Bigr)                   \nonumber\\
    && +\frac{1}{2}\partial_r\Bigl( \chi^A\chi^B \eta_5^{C}\partial_\theta(w_5 \cos\theta + w_4)- \chi^A\chi^B (\eta_5^C \cos\theta  \nonumber\\
    && + \eta_4^C)\partial_\theta w_5 +2 \chi^A(\eta_5^B\partial_\theta \eta_4^C - \eta_5^B \eta_5^C \sin\theta - \partial_\theta \eta_5^B \eta_4^C          )       \Bigr)  \nonumber\\
    && - \frac{1}{2}\partial_\theta \Bigl( \chi^A\chi^B \eta_5^{C}\partial_r(w_5 \cos\theta + w_4)- \chi^A\chi^B (\eta_5^C \cos\theta  \nonumber\\
    && + \eta_4^C)\partial_r w_5 +2 \chi^A(\eta_5^B\partial_r \eta_4^C  - \partial_r \eta_5^B \eta_4^C  )                   \Bigr) \Bigr] \nonumber\\
    && \times  dt\wedge dr\wedge d\theta \wedge d\phi \wedge d\psi \;.
\end{eqnarray}

When defining
\begin{equation}
    \label{psitilde} \chi^A =  - s f \,  X^A \,,
\end{equation}
the terms in (\ref{app1})
\begin{equation} \label{eq:1}
\begin{split}
&\frac{1}{3V}  C_{ABC} \Bigl[+3 \chi^A \Bigl( \partial_r(\eta_5^B\cos\theta + \eta_4^B)\partial_\theta \eta_5^C -\partial_\theta(\eta_5^B\cos\theta + \eta_4^B)\partial_r \eta_5^C  \Bigr)  \Big]             \\
 &= (2 G_{AB} \, s f - 9 \frac{X_A X_B}{V^2}s f  ) \Bigl( \partial_r(\eta_5^B\cos\theta + \eta_4^B)\partial_\theta \eta_5^A -\partial_\theta(\eta_5^B\cos\theta + \eta_4^B)\partial_r
 \eta_5^A
 \Bigr).
 \end{split}
 \end{equation}

\end{document}